# Beltrami Flows, Non-Diffracting Waves and the Axion Beltrami-Maxwell Postulates

THEOPHANES E. RAPTIS[1,2], CHRISTOS D. PAPAGEORGIOU[3]

[1]Division of Applied Technologies,
National Center for Science and Research "Demokritos",
Patriarchou Gregoriou & Neapoleos, Athens, Greece.

[2]Informatics and Telecommunications Dept.,
University of Peloponnese, Tripolis, Greece.

[3]Dept. of Electrical & Electronic Engineering
National Technical University of Athens
Zografou Campus, 9 Iroon Polytechniou Str. 15780
GREECE
chrpapag@gmail.com

*Abstract:* - We present a particular class of solutions in Cartesian, cylindrical and spherical coordinates of the non-dispersive travelling wave variety that propagate an envelope of varying vorticity some of which include topological waves with parallel electric and magnetic components. The significance of these solutions is examined in the recently proposed Axion-Maxwell field theory with potential applications in material science and topological insulators.

## 1 Introduction

By the end of the previous century, Maxwell theory got some important influences from other related fields like hydrodynamics and plasma physics which enriched the existing framework. This has already had a precedent in the work of Maxwell himself [1], [2] who borrowed ideas from ideal Euler hydrodynamics and similar fields since at that time, the conception of the field abstraction was that of a material entity or a "fluid" structure tied up with the aether concept.

Even after the collapse of the aether model by relativistic theories, the many similarities of Maxwell theory with hydrodynamic models were often revived as in Marmanis [3], Holland [4], and Kambe [5]. More recent applications of Maxwell theory in complex phenomena like plasma dynamics, heliodynamics and magneto-hydrodynamics are reviewed in a monography by Marsh [6] where the role of the so called "*force-free*" fields is raised in association with the work of Eugenio Beltrami [7] in ideal hydrodynamics. Beltrami flows are of a characteristic topology akin to that of tornado flows and correspond to the "eigenfields" of the *curl* operator.

These ideas found later use by Lakhtakia [8] in the form of the now known as the Beltrami-Maxwell Postulates where all cases of the electromagnetic field are reducible into generalized, complex Beltrami flows. These also naturally allow non-abelian generalizations with magnetic monopole currents. A restricted version of the general transform has been also introduced previously in certain versions of covariant electrodynamics with the so called Riemann-Silberstein vector [9], [10] as a complex combination of the *E* and *B* components of free space solutions which can be traced back to the symmetrization of the electromagnetic tensor. The notion has become again popular in recent reformulations of electromagnetism based on geometric algebra [11].

One of the strangest varieties of self-dual fields strongly associated with Beltrami flows has been brought to light in the original work by Yabe and Mushiake [12] as well as Chu and Ohkawa [13], raising some controversy due to their parallel electric and magnetic components. Later, these were suggested by Gray [14] as special solutions of Maxwell equations while Shimoda, Kawai and Uehara proceeded in a partial recovery of both stationary and travelling waves with similar characteristics [15]. The controversy was only recently settled with a complete proof and classification from Nishiyama [16] based on previous work by Ko and Jang [17] as well as Low [18] on the complete solutions of Beltrami equation revealing both plane and spherical travelling waves with parallel electric and magnetic components.

While such waves may carry energy but no momentum, they can be understood as topological excitations propagating geometric information of the field structure, similar to certain magnetic recombination processes in plasma physics or certain varieties of topological solitons [19]. In some controlled environments with appropriate materials they could even serve as extreme computational elements as in bounded domains where tweaking the information contained in appropriately specified boundaries would demand a rearrangement of the internal energy and the field structure.

Axion theory on the other hand has being born out of an attempt to resolve symmetry problems of quantum chromodynamics (QCD) and the prediction of CP violations that were never observed. A possible resolution was given in the context of Perccei-Quinn theory [20] which adds a symmetry breaking *θ-term* in the QCD Lagrangian leading to a new, hypothetical elementary particle via its promotion to a quantized pseudoscalar field, first proposed by Weinberg [21] and Wilczek [22]. As of now, any strong violations have been excluded from measurements of the neutron electron dipole moment leaving only the possibility for an extremely small value of the *θ* parameter and a mass less than $10^{-11}$ of that of the electron [23]. Due to this, the interest in the theory spiked recently out of the possibility of axions being the true components of cold dark matter via a boson condensation mechanism [24] since there should be fairly stable, with even lower states of mass-energy being extremely improbable.

It was recently shown by a Chinese group [25] that an equivalent Axion field can appear in the form of magnetic fluctuations coupled to the electromagnetic field in topological insulators enabling a nonlinear modulation of the EM field. Additionally, in a following work, Vissineli [26] examines the consequences of the interaction of an Axion field with the classical EM field based on the symmetry of the standard electromagnetic duality via a coupling mechanism that only affects Gauss and Faraday laws. This then results in a new, complete set of five equations by adding to the modified Maxwell equations an inhomogeneous Klein-Gordon equation for the pseudoscalar massive term of the additional Axion field.

In what follows, we utilize a unique class of Beltrami flows for the vector potential itself which naturally contains a class of free space travelling waves with parallel electric and magnetic components. We first examine general characteristics of the new class in section *2*, while in section *3*, we seek the most general solutions along the lines laid in [16]. In section *4*, we examine the conditions for the previously found types of waves to exist in the modified Axionic Maxwell equations and propose a hypothesis for the extremely weak mass values while, in section *5*, we conclude.

## 2 Problem Formulation

We start by finding special solutions of the generic problem for the vector potential corresponding to a Beltrami flow defined by

$$\nabla \times \mathbf{A} = \lambda(\mathbf{r},t)\mathbf{A} \qquad (1a)$$

$$\nabla\lambda \cdot \mathbf{A} = \nabla \cdot \mathbf{A} = 0 \qquad (1b)$$

Eq. (1a) defines a proper Beltrami flow while the first of the *lhs* in eq, (1b) assures consistency with the fundamental vectorial identity $\nabla \cdot (\nabla \times \mathbf{A}) = 0$. The problem defined by equations (1)-(3) has been originally tackled for special cases of constant $\lambda$ ("*Trkal fields*") during the second half of the previous century with the work of Chandrasekhar and Kendal [27], Woltjer [27], Moses[29] and others in the case of astrophysical magnetic fields which prefer force-free states for minimizing the magnetic part of the Lagrangian due to elimination of the magnetic part of the Lorentz force after alignment of the current with the magnetic field.

In our case we are interested in a particular version of the general problem applied directly to the vector potential in the Coulomb gauge (eq. (2b)). The advantage of using (1a) is that it leads to certain types of sources that can allow modulating both the vorticity of the resulting fields as well as their total angular momentum content. We then use the fact that for non-constant $\lambda$ scalars the same solutions will necessarily satisfy the general D'Alambertian operator $[] = \nabla^2 - c^{-2}\partial_t^2$ with sources as can be seen by direct application of a second curl operator giving $\nabla \times \nabla \times \mathbf{A} = -\nabla^2 \mathbf{A}$.

Then the total D'Alambertian becomes

$$[] \cdot \mathbf{A} = -\nabla \times (\lambda \mathbf{A}) - c^{-2}\partial_t^2 \mathbf{A} \quad (2)$$

From standard vector identities we have then the immediate general result

$$[] \cdot \mathbf{A} = -\lambda^2 \mathbf{A} - \nabla \lambda \times \mathbf{A} - c^{-2}\partial_t^2 \mathbf{A} = \mu_0 \mathbf{J} \quad (3)$$

In (3) *J* stands for an equivalent source current that if present in the standard Maxwell equations with linear constitutive relations, than it would simultaneously cause the appearance of the standard fields defined as

$$\mathbf{E} = -\nabla \phi - \partial_t \mathbf{A}, \quad \mathbf{B} = \nabla \times \mathbf{A} = \lambda \mathbf{A} \quad (4)$$

Checking the *rhs* with the combined use of (1b) and the vector identity

$$\nabla \times (\nabla \lambda \times \mathbf{A}) = \nabla \lambda (\nabla \times \mathbf{A}) = \nabla (|\lambda|^2 / 2) \mathbf{A},$$

we find that it is divergence-free and hence this is a purely transverse (solenoidal) source. Since such a current does not require the presence of any static charge we can drop $\varphi$ entirely. We notice that in this case, if a certain discrete symmetry is present, a linear superposition of all group members will fall again into the special category of electromagnetic fields with parallel electric and magnetic components. Examples can be given with the aid of general solutions discussed in the next section.

The above source can be written in the form of a linear transform of the vector potential utilizing the tensorial representation of the cross product through the linear operator

$$\hat{\mathbf{R}} = [\nabla \lambda \times] = \begin{pmatrix} 0 & -\partial_3 \lambda & \partial_2 \lambda \\ \partial_3 \lambda & 0 & -\partial_1 \lambda \\ -\partial_2 \lambda & \partial_1 \lambda & 0 \end{pmatrix}$$

In the above $\partial_i$ stands for the elements of the gradient 3-vector with the appropriate metric elements in the chosen coordinate system. Then the sources can be expressed in the operator matrix representation as

$$\mu_0 \mathbf{J} = -\hat{\mathbf{K}} \cdot \mathbf{A} - c^{-2}\partial_t^2 \mathbf{A}, \quad \hat{\mathbf{K}} = \hat{\mathbf{R}} + \lambda^2 \hat{\mathbf{I}} \quad (5)$$

In the next section we seek alternative, non-harmonic solutions of (1a-b) by introducing an ansatz for both the vector potential as well as the eigen-vorticity $\lambda$, to be functions of the single variable $\mathbf{u} = \mathbf{r} - Vt$ where *V* an arbitrary velocity such that $c^{-1}\partial_t \mathbf{A} = -\beta \partial_\mathbf{u} \mathbf{A}$ and $c^{-2}\partial_t^2 \mathbf{A} = \beta^2 \partial_\mathbf{u}^2 \mathbf{A}$ where $\beta = V/c$ is the usual relativistic notation. Notably, there are appropriate choices of the transverse part of *A* such that the cross product in (3) and (5) can be made to vanish so that we can guarantee the $E \sim A$ structure of the fields.

Such waves are often members of the class of non-dispersive or, non-diffracting solutions of Maxwell equations which also satisfy the additional linear equation

$$(\nabla^2 - V^{-2}\partial_t^2)\mathbf{A} = []_V \cdot \mathbf{A} = 0 \quad (6)$$

The class of non-diffracting waves has been well studied for three decades after the original finding of Focused Wave Modes by Brittingham. There are by now a variety of known and practically applicable cases of such solutions of the linear wave equation either in the field of paraxial optics or linear acoustics known as Bessel, Gaussian or Airy beams, as well as X-waves first proposed in the work of Brittingham [30], Hillion [31] and later verified by Lu and Greenleaf [32]. The field is nicely reviewed in a recent monography [33]. In most of these cases the resulting field structures correspond to localized wavepackets.

Since we also demand the simultaneous satisfaction of the original equation $[]_c \cdot \mathbf{A} = \mathbf{J}$ for them to also satisfy Maxwell equations, and noticing that $V^{-2}\partial_t^2 \mathbf{A} = \partial_\mathbf{u}^2 \mathbf{A}$ we may immediately derive a constraint on the current sources using the additional relativistic $\gamma$ factor as

$$\mathbf{J} = \frac{1}{\mu_0}(\beta^2 - 1)\partial_\mathbf{u} \mathbf{A} = -\left(\frac{1}{\mu_0 \gamma^2}\right)\partial_\mathbf{u} \mathbf{A} \quad (7)$$

Using (5) we may also write the additional condition for the orthogonal projection

$$\nabla \lambda \times \mathbf{A} = (2\beta^2 - 1)\partial_u \mathbf{A} - \lambda^2 \mathbf{A} \quad (8)$$

It is then straightforward to see from (7) that free space solutions exist exactly at the pole $\beta = \pm 1$ or $V = \pm c$ where any sources disappear. By a curious coincidence, we also notice that such a form of current in the case where $\partial_u \mathbf{A} \sim \mathbf{A}$ that will be examined in next sections, also appeared previously in the extended Proca electrodynamics with massive photons [34] [35] via the correspondence $\lambda(\mu_0 \gamma^2)^{-1} \to m_{ph}^2$. The authors do not know at the moment whether such a coincidence could be meaningful in some alternative framework, nor is it the aim of the present report although mention of is given at the end of section *4*. We next examine specific solutions of (1a-b) with the required properties before we explore their possible significance in the case of the modified Maxwell equations coupled with an additional Axion field.

## 3 Explicit general solutions of Beltrami conditions

We examine the conditioned equation (1a-b) in Cartesian, cylindrical and spherical coordinates. The conditions in (1b) can be automatically satisfied with the choice of $u = z - Vt$ for the restricted class of vector fields $A = [f(u), g(u), 0]$. In the simplest case of a Cartesian system one has a 2-ODE system as

$$\begin{cases} \partial_z g(u) + \lambda(u) f(u) = 0 \\ \partial_z f(u) - \lambda(u) g(u) = 0 \end{cases} \quad (9)$$

The above evidently has a Hamiltonian symmetry which can be emphasized by rewriting it in the form $\partial_z \mathbf{A}_\perp = \mathbf{D} \cdot (\lambda \mathbf{A}_\perp)$ where $\mathbf{D}$ is the Darboux symplectic matrix $\begin{pmatrix} 0 & 1 \\ -1 & 0 \end{pmatrix}$ and $\mathbf{A}_\perp = (f, g)$. We notice that there is a scalar function $H$ such that $\nabla_{f,g} H \cong \lambda \mathbf{Y}$ which is just a modulated harmonic oscillator $H = (f^2 + g^2)(\lambda/2)$ with an associated pair of modulated creation-annihilation operators as $\sqrt{\lambda/2}(\alpha, \alpha^+)$. A propagator for the system (9) is directly given as $\exp(-\lambda \mathbf{D} u)$ and a general solution vector can be given as a superposition of two Hertzian potentials in the form

$$\mathbf{A} = \sin(w(u))\mathbf{r}_0 + \cos(w(u))\mathbf{r}_1, \quad \lambda(u) = \frac{dw}{dz} \quad (10)$$

Here, $\mathbf{r}_0 = c_1 \mathbf{e}_x + c_2 \mathbf{e}_y$ and $\mathbf{r}_1 = \mathbf{D} \cdot \mathbf{r}_0 = c_2 \mathbf{e}_x - c_1 \mathbf{e}_y$ with $c_{1,2}$ arbitrary constants where, and $w$ is any regular analytic function which serves as a superpotential. The resulting electric and magnetic components will then satisfy $\mathbf{E} = \mathbf{D} \cdot \mathbf{B} = \lambda(\mathbf{D} \cdot \mathbf{A})$ for a generic source current $\mathbf{J} = -(\lambda/\mu_0 \gamma^2)\mathbf{A}$ while free space solutions are obtained with $u = z - ct$. Localized solutions for particular choices of the modulating $w$ factor may stand for a type of EM bullet. We notice that the action of the *curl* operator on this class of fields can have an algebraic correspondence with $\mathbf{D}$ as $\mathbf{D}_z^\partial = \begin{pmatrix} 0 & \partial_z \\ -\partial_z & 0 \end{pmatrix}$. This symmetry allows producing a new solution vector simply by taking $\mathbf{A}' = \mathbf{D} \cdot \mathbf{A}$ and since $\mathbf{D}$ is anti-involutive this leads to $\mathbf{D} \cdot \mathbf{E}' = \mathbf{B}' = \lambda \mathbf{A}'$. A superposition of these two fields will generate a total field with parallel electric and magnetic components. With similar rotations one can also produce dual combinations as $\mathbf{M} \cdot \mathbf{E} = -\mathbf{M} \cdot \mathbf{B} = -\lambda(\mathbf{M} \cdot \mathbf{A})$ where $\mathbf{M} = \mathbf{I} + \mathbf{D}$.

As a matter of fact, one can also turn this into an algebraic symmetry by taking any arbitrary pair of multiplicative inverses such that $(\partial_z f)g = -(\partial_z g)f$. If we let then $u = f - g$, $v = f + g$, we naturally have that $\pm \partial_z u = -\partial_z(\log g)(\pm v)$, $\pm \partial_z v = -\partial_z(\log g)(\pm u)$ so that the combined action of $\mathbf{D}_z^\partial$ on any of the four vector fields $\mathbf{F} = \pm u \mathbf{e}_x \pm v \mathbf{e}_y$ is to map them into $-\lambda \mathbf{F}$ with $\lambda = \partial_z(\log g)$.

To locate more general solutions we follow the method of Nishiyama, detailed in [16] where we find at least three classes of waves, the first two being stationary waves while Type III solutions allow for non-transverse travelling waves. Following the notation in [16], general Cartesian solutions can be found from an extension of (10) where one obtains

$$f - \mathbf{i}g = \Psi(x + \mathbf{i}y)e^{\mathbf{i}G(u)}, \quad \lambda = dG/du \quad (11)$$

Here, $\Psi$ is any holomorphic function satisfying the Cauchy-Riemann conditions for its real and imaginary parts. Rewriting (11) in cylindrical coordinates we get

$$A_\rho - \mathbf{i}A_\varphi = (f - \mathbf{i}g)e^{\mathbf{i}\varphi} = \Psi(\rho e^{\mathbf{i}\varphi})e^{\mathbf{i}G(u)+\mathbf{i}\varphi} \quad (12)$$

The solution in (10) is indeed recoverable from (12) for the choice $\Psi(\rho e^{\mathbf{i}\varphi}) = (\rho e^{\mathbf{i}\varphi})^{-1}$ plus a symmetric superposition. We shall refer to these types as the "*z-waves*".

For the cylindrical case a general discussion is given in Ko and Jiang [] where the general solution for the case $\mathbf{A}(\rho,\varphi,\zeta,t)$ presents a particular mathematical difficulty that has not being tackled yet. Still, it is possible to restrict the class of solutions to the case $\mathbf{A}(\rho,u), u = \rho - vt$ where their general result can be given as an expansion over arbitrary real constants $K_n$ and arbitrary, at least $C^1$ differentiable functions $G_n$ of the form

$$\begin{aligned}\mathbf{A} &= A_\varphi \mathbf{e}_\varphi + A_z \mathbf{e}_z \\ A_\varphi^2 &= \sum_n K_n^2 \cos^2(G(u))e^{-2Y} \\ A_z^2 &= \sum_n K_n^2 \sin^2(G(u))e^{-2Y}\end{aligned} \quad (13)$$

In (13), we have

$$\lambda = -\frac{1}{A_\varphi}\frac{dA_z}{d\rho}, \quad Y = \int \frac{d\rho}{\rho}\cos^2(G_n(u)) \quad (14)$$

Particularly, setting $K_n = 0$ except for $n = 1$, we obtain

$$\begin{aligned}\mathbf{A} &= e^{-\int \frac{d\rho}{\rho}\cos^2(G_n(u))}\mathbf{F} \\ \mathbf{F} &= \cos(G(u))\mathbf{e}_\varphi - \sin(G(u))\mathbf{e}_z\end{aligned} \quad (15)$$

For this case we also have

$$\lambda(\rho,u) = dG/du - \mathrm{sin}c(G(u)) \quad (16)$$

We shall also refer to these waves as "*ρ-waves*".

The spherical case from the generic Chang–Carovillano–Low results in [16] can be written as

$$A_r = 0, \quad A_r - \mathbf{i}A_\varphi = \frac{\Psi(y+\mathbf{i}\varphi)}{r\sin\varphi}e^{\mathbf{i}G(u)}, \quad \lambda = dG/du \quad (17)$$

Here, we employ a periodic holomorphic function such that $\Psi(c) = \Psi(c + 2\pi\mathbf{i}), c \in \mathbf{C}$ with the real argument being a function of the first spherical angle as $y = \log(\tan(\theta/2))$. Periodicity implies an equivalent expansion with complex constants $C_m$ as

$$\begin{aligned}\Psi(c) &= \sum_{m=-\infty}^{\infty} C_m e^{m(y+\mathbf{i}\varphi)} \\ &= \sum_{m=-\infty}^{\infty} C_m e^{\mathbf{i}m\varphi}\left(\tan\frac{\theta}{2}\right)^m\end{aligned} \quad (18)$$

A particularly simple case comes out of the choice $\Psi = 1$ which leads to

$$\mathbf{A} = \frac{1}{r\sin\varphi}\left(\cos(G(u))\mathbf{e}_\theta - \sin(G(u))\mathbf{e}_\varphi\right) \quad (19)$$

We shall refer to these waves as the "*r-waves*".

Last but not least, we would like to stress the fact that many members of the above three classes may not always be physically realizable due to their singular Fourier spectrum, hence the choice of the generic holomorphic generators $\Xi$, $Y$ or $G$ will have to be constrained by the demand of a finite spectrum which further narrows down the class of physically acceptable solutions. For cases of localized waves and especially in the Cartesian case of *z*-waves this is easier to accomplice. Having exhausted the possible set of analytically known solutions, we may now proceed in examining their possible significance in the case of the Axionic extensions of Maxwell equations.

## 4 Beltrami flows in Axion electrodynamics

The general duality symmetry of Maxwell equations with or without sources is defined from the invariance under an SO(2) matrix operator $\mathbf{R}$ applied to a pair of fields $(\mathbf{E},\mathbf{B}) \rightarrow (\mathbf{E}',\mathbf{B}') = \mathbf{R}\cdot(\mathbf{E},\mathbf{B})^T$ where the new pair is again another valid solution. Introduction of an additional pseudo-scalar field $\theta(\mathbf{r})$ causes a symmetry breaking via an additional term in the electromagnetic Lagrangian

$$L_0 = (1/2)\left(c^{-1}|\mathbf{E}|^2 - |\mathbf{B}|^2\right) + \rho\phi + \mathbf{J}\cdot\mathbf{A}$$

as $L_0 \to L_0 + (\kappa/\mu_0 c)\theta(\mathbf{r},t)(\mathbf{E}\cdot\mathbf{B}) + L_A$. The additional pure Axion Lagrangian also includes a new potential as $L_A = (1/2)\partial^\mu\theta\partial_\mu\theta + U(\theta)$.

In [24], a new modified version of Maxwell equations which admit dynamic duality transformations is given with magnetic monopole charges and currents and an additional degree of freedom for the pseudo-scalar field $\theta$ satisfying an inhomogeneous Klein-Gordon operator with a new current term as

$$j_\theta = -(\kappa\theta/\mu_0 c)(\mathbf{E}\cdot\mathbf{B}) + \partial_\theta U \qquad (20)$$

This time, the duality rotation adopts a new degree of freedom via the correspondence of the internal angle $\xi$ with the new field $\theta$ as $\tan\xi = -\kappa\theta$ and after performing a rotation that makes monopole charges and currents to vanish as in [24] we have

$$\begin{pmatrix}\mathbf{E}'\\\mathbf{B}'\end{pmatrix} = \begin{pmatrix}1 & -c\kappa\theta\\c\kappa\theta & 1\end{pmatrix}\begin{pmatrix}\mathbf{E}\\\mathbf{B}\end{pmatrix} \qquad (21)$$

The five augmented Axionic Maxwell equations for linear media then become

$$\nabla\cdot\mathbf{E}' = \rho/\varepsilon_0$$
$$\nabla\cdot\mathbf{B}' = \mu_0\rho_m$$
$$\nabla\times\mathbf{E}' = -\partial_t\mathbf{B}' \qquad (22\text{a-e})$$
$$\nabla\times\mathbf{B}' = c^{-2}\partial_t\mathbf{E}' + \mu_0\mathbf{J}_e$$
$$[\,]\theta = -\delta\mathbf{E}\cdot\mathbf{B} - \partial_\theta U$$

where $\delta = \kappa/\mu_0 c \sim 0.002\kappa$. The cost of making the rotation dynamic is to alter the gauge freedom and additional gauges have been introduced to that purpose so that one obtains an axionic electrodynamics with three wave equations. After performing an internal rotation removing the magnetic sources the two revised electromagnetic wave equations with the new gauges become

$$[\,]\phi' = \rho_e + \partial_t S, \quad [\,]\cdot\mathbf{A}' = \mathbf{J}_e - \nabla S \qquad (23a)$$

$$[\,]\phi = \rho_e + \kappa\nabla\theta\cdot\mathbf{B} + \partial_t S$$
$$[\,]\cdot\mathbf{A} = \mathbf{J}_e - \kappa(\partial_t\theta\mathbf{B} + \nabla\theta\times\mathbf{B}) - \nabla S \qquad (23b)$$

Here, $S = \nabla\cdot\mathbf{A} + \partial_t\phi$, comes from the standard part of the Lorentz gauge.

We next observe that the original map in (19) is formally similar with the transform proposed by Lakhtakia [8] in the so called, Beltrami-Maxwell reformulation with $-\kappa\theta \to \mathbf{i}Z$ where $Z = \sqrt{\mu_0/\varepsilon_0}$. The same can also be applied in the case of (20) as a composite map to produce the axionic analogue of complex Beltrami-Maxwell fields as

$$\begin{pmatrix}\mathbf{Q}_+\\\mathbf{Q}_-\end{pmatrix} = (\mathbf{I} + (\mathbf{i}Z)\mathbf{D})\cdot(\hat{\mathbf{I}} + (\kappa\theta)\mathbf{D})\begin{pmatrix}\mathbf{E}\\\mathbf{B}\end{pmatrix}$$
$$= (\alpha\mathbf{I} + \beta\mathbf{D})\begin{pmatrix}\mathbf{E}\\\mathbf{B}\end{pmatrix} \qquad (24)$$

Here, $\alpha = \mathbf{i}Z\kappa\theta$, $\beta = \kappa\theta - \mathbf{i}Z$. Given also new sources defined as $s_\pm = (1/2)(-\rho_m \pm \mathbf{i}Z\rho_e)$, $\mathbf{S}_\pm = s = (1/2)(-\mathbf{J}_m \pm \mathbf{i}Z\mathbf{J}_e)$ the new fields will then satisfy

$$\nabla\cdot\mathbf{Q}_\pm = \pm\mathbf{i}cs_\pm$$
$$\nabla\times\mathbf{Q}_\pm = \pm\mathbf{i}c^{-1}\partial_t\mathbf{Q}_\pm + \mathbf{S}_\pm$$
$$[\,]\theta = -\delta\left(c\mathbf{Q}_+\cdot\mathbf{Q}_- - d|\mathbf{Q}_+|^2 - d|\mathbf{Q}_-|^2\right) - \partial_\theta U$$
$$c = Z^2 - Z\kappa\theta(Z(\kappa\theta - \mathbf{i}) + \kappa\theta)$$
(25a-d)

Let then $(\hat{T})(\mathbf{E},\mathbf{B}) \to (\mathbf{E}',\mathbf{B}')$ be the action of (21) on some abstract bivector to a new one in the space of solutions of (22). We want to examine the field structure entailed by this map as an endomorphism in the set of all vector fields admissible by the new Axion-Maxwell postulates. The eigenspace analysis of (19) gives a set of eigenvalues and eigenvectors as

$$l_\pm = |l_\pm|^{1/2}e^{\pm\mathbf{i}\xi}, \mathbf{e}_{1,2} = |l_\pm|^{-1/2}[1, \pm c\kappa\theta]$$
$$|l_\pm|^2 = 1 + (c\kappa\theta)^2 \qquad (26)$$

Let then $\mathbf{A}'$ a vector potential such that (22) are satisfied by a bivector with elements $\mathbf{e}' = -(1/c\kappa\theta)\mathbf{b}'$ where $\kappa$ an arbitrary scalar factor such that $(\mathbf{e}',\mathbf{b}') \propto (1,-c\kappa\theta)\mathbf{e}$, and $(\hat{T}^{-1})(\mathbf{e}', \mathbf{b}') \rightarrow (\mathbf{e}, \mathbf{b}) = l_-^{-1}(\mathbf{e}', \mathbf{b}')$. We conclude that strictly parallel electric and magnetic components are not in general possible without further restricting the choice of the vorticity and the resulting proportionality factor in terms of some solution of the last equation in 22(a-e).

Since equations (22a-e) are identical with the standard Maxwell system in the absence of monopoles one expects the existence of free space solutions as those in section *3* as well as general non-diffractive solutions for non-zero solenoidal current sources. The original fields will then have the general form

$$\mathbf{E} = -\frac{\beta(\partial_\mathbf{u} \mathbf{A}') - (\kappa\theta\lambda)\mathbf{A}'}{1+(\kappa\theta)^2}$$
$$\mathbf{B} = -\frac{\beta(\partial_\mathbf{u} \mathbf{A}') + (\kappa\theta\lambda)\mathbf{A}'}{1+(\kappa\theta)^2} \quad (27)$$

We then notice that the new gauge conditions of (23a-b), partially fix the choice of the arbitrary vorticity scalar *λ* as a function of *θ*. Indeed, for the consistency of the transformed fields with the additional gauge terms all charges must vanish in the transformed fields generated by $\mathbf{A}'$ and for this we must also have $\nabla\theta \cdot (\lambda\mathbf{A}') = 0$. Assume then a generic dependence as *λ(θ(u))* where *u = x − vt* in which case, condition (1b) guarantees that $\nabla\lambda \cdot \mathbf{A}' = (d\lambda/d\theta)\nabla\theta \cdot \mathbf{A}' = 0$. In case of parallel electric and magnetic components $\mathbf{E}' = -\mu\mathbf{B}' = \lambda\mathbf{A}$, (27) simplifies as

$$\mathbf{E} = \varphi_-(\theta)\lambda(\theta)\mathbf{A}'$$
$$\mathbf{B} = \varphi_+(\theta)\lambda(\theta)\mathbf{A}' \quad (28)$$

where, $\varphi_-(\theta) = (c\kappa\theta - \mu)(1+(\kappa\theta)^2)^{-1}$ and $\varphi_-(\theta) = (c\kappa\theta\mu + 1)(1+(\kappa\theta)^2)^{-1}$.

For free space solutions, one should also assure that the gauge current term $(\partial_t\theta)\mathbf{B} + \nabla\theta \times \mathbf{B}$, also vanishes. We can then utilize relation (8) of section *2*, to get $\nabla\theta \times (\lambda\mathbf{A}') = (d\lambda/d\theta)^{-1}(\partial_u - \lambda^2)\mathbf{A}'$. This then allows eliminating the 2nd derivative from the *θ* wave equation by an additional time differentiation over the constraint of the vanishing current in which case we get

$$c\lambda(d\lambda/d\theta)(\partial_t\theta)\mathbf{A}' = (\partial_u - \lambda^2)\mathbf{A}' \quad (29)$$

Since all derivatives can now be eliminated we are left with a functional equation relating *λ(θ)* with *θ*. The situation is further simplified in the case of fields for which $\partial_u \mathbf{A} \propto \mathbf{A}$.

Last but not least, the recent strong constraints on the extremely small Axionic mass could have some relevance in the case of sub-luminal formations of large scale non-diffractive travelling waves from weak, non-relativistic slow current sources of galactic origin. In particular taking into account the observations at the end of section *2*, and adopting the convention of an harmonic axion potential as $U(\theta) \sim (m\theta)^2/2$ we notice that both the vector potential and the *θ* field will satisfy a pair of similar Klein-Gordon operators as

$$\left([\,] - M_e^2\right) \cdot \mathbf{A}' = 0 \quad (30a)$$

$$\left([\,] - m^2\right)\theta = \delta\phi_-(\theta)\phi_+(\theta)\lambda^2(\theta)|\mathbf{A}'|^2 \quad (30b)$$

Here, we use the effective mass term as $M_e^2 \approx -\lambda(\theta)/\mu_0$ assuming parallel electric and magnetic components of a sub-luminal cyclonic vorticity following a weak ionic current. The *rhs* of (28b) contains some hard nonlinearity but since there is no unique choice for *λ* we can make a reasonable choice at least in 1st order as $\phi_-(\theta)\phi_+(\theta)\lambda^2(\theta) \approx k_0 - k_1\theta + ...$ with *k* a constant. Then a simplified equation is obtained as

$$\left([\,] - m^2 + k_1\delta|\mathbf{A}'|^2\right)\theta \approx k_0|\mathbf{A}'|^2 \quad (31)$$

The peculiar appearance of a new effective mass as $m_e^2 = m^2 - k\delta|\mathbf{A}'|^2$ suggests here a kind of oscillatory "screening" effect which although weak could have an additional contribution while existing data for equipartitioned magnetic fields strength provide an order of magnitude as $10^{-9}$ Tesla. While such an argument appears speculative for the original cold dark matter proposal, it could still be of some importance in applications like topological insulators in limited, controlled environments.

## 5 Conclusions

We discussed the unique case of some recently found special solutions of Maxwell equations and in particular a special class of them having the form of non-diffracting travelling waves with both transverse and parallel electric and magnetic components. We then extracted some general expressions for this class based on previous mathematical analysis of the resulting equations for the vector potential originating in the classification of vector fields in ideal hydrodynamics by Beltrami. The resulting electromagnetic formations seem to resemble a sort of moving "vorticity walls" or self-focused wave modes that could easily be passed as an unexpected kind of longtitudinal radiation in the absence of a phased array detector for determination of the vorticity components. We noticed the particular significance that these modes could have in the case of the proposed augmentation of electrodynamics in axion field theory and we examined their influence on the additional axion field equation showing the possibility of an oscillating axion mass.